\documentclass[a4paper,amsmath,amssymb]{jpconf}

\usepackage{epsfig,graphicx,amsmath, amsthm, amssymb}
\usepackage[mathscr]{euscript}

\newcommand{\pprime}{{\prime\prime}}

\newcommand{\bra}{\langle}
\newcommand{\ket}{\rangle}

\newcommand{\order}{{\mathcal O}}

\newcommand{\one}{{\rm 1\!\!I}}

\newcommand{\be}{\begin{equation}}
\newcommand{\ee}{\end{equation}}
\newcommand{\bd}{\begin{displaymath}}
\newcommand{\ed}{\end{displaymath}}

\newcommand{\R}{{\rm I\!R}}

\newcommand{\bc}{\ensuremath{\mathbf{c}}}
\renewcommand{\be}{\ensuremath{\mathbf{e}}}

\newcommand{\bM}{\ensuremath{\mathbf{M}}}

\newcommand{\bpsi}{{\mbox{\boldmath $\psi$}}}
\newcommand{\bphi}{{\mbox{\boldmath $\phi$}}}

\newcommand{\bsigma}{{\mbox{\boldmath $\sigma$}}}

\newcommand{\Prob}{\mathscr{P}}

\newcommand{\A}{\mathscr{A}}
\newcommand{\CC}{\mathscr{C}}

\newcommand{\FF}{\mathscr{F}}
\newcommand{\BB}{\mathscr{B}}
\newcommand{\WW}{\mathscr{W}}

\newcommand{\atan}{{\rm atan}}

\newcommand{\fs}{\footnotesize }

\begin{document}

\title[Replica methods for loopy sparse random graphs]{Replica methods for loopy sparse random graphs$^\dag$}
\author{ACC Coolen}

\address{Department of Mathematics,  
King's College London, \\
The Strand, 
London
WC2R 2LS, United Kingdom
}

\ead{ton.coolen@kcl.ac.uk}



\begin{abstract}
I report on the development of a novel statistical mechanical formalism for the analysis of random graphs with many short loops, and processes on such graphs. The graphs are defined via maximum entropy ensembles, in which both the degrees (via hard constraints) and the adjacency matrix spectrum (via a soft constraint) are prescribed. The sum over graphs can be done  analytically, using a replica formalism with complex replica dimensions. All known results for tree-like graphs are recovered in a suitable limit. For loopy graphs, the emerging theory has an appealing and intuitive structure, suggests how message passing algorithms should be adapted, and what is the structure of theories describing spin systems on loopy architectures. However, the formalism is still largely untested, and may require further adjustment and refinement. 
\end{abstract}

\vspace*{-0mm} 


\section{Introduction}

Networks and graphs are increasingly popular and effective tools for visualising and modelling large and complex processes and `big' data sets. We know that many important graphical structures in the world (biological networks, computing and communication networks, resource grids, lattices in physics, etc) are not tree-like; they tend to have many short loops, and we know that processes on graphs are affected significantly by the presence of such loops. It is therefore problematic that most of our tools and algorithms for analysing (processes on) finitely connected graphs, such as cavity methods \cite{MezardParisi2001}, belief propagation type algorithms \cite{belief1,belief2,montanari}, and conventional replica analyses \cite{CoolenPerez,guzaipaper}, require topologies that are locally tree-like. Some methods were extended with loop corrections \cite{MontanariRizzo,ParisiSlanina,ChertkovChernyak,MooijKappen}, but all tend to fail for graphs with many short loops. The exceptions,  solvable spin models or spectrum calculations for loopy graphs,  all rely on some special property of either the dynamics or the graph topology, and are thus  non-generic: these include spherical models \cite{BerlinKac}, one- and two-dimensional Ising systems \cite{Onsager,Baxter}, loopy immune models that can be mapped to tree-like systems \cite{immune_model}, and trees of loopy modules \cite{MozeikaCoolen,Metz_etal}.

In this paper I report on ongoing research aimed at the development of a new and more general statistical mechanical method that removes the restriction to tree-like graphs. It is designed to handle analytically ensembles of large and sparse random graphs with prescribed degree sequences and prescribed loop statistics (via their adjacency spectra), and stochastic processes on such graphs. It is based on an alternative flavour of the replica method, with imaginary replica dimensions, and produces explicit closed equations in the infinite size limit, leading to Shannon entropies and expressions for spectra of ensembles of sparse loopy graphs. The familiar equations describing tree-like graphs are recovered as a simple limiting case. 
\vspace*{\fill}

{\noindent\fs
$\dag$ This paper is dedicated to the memory of our colleague and friend Jun-Ichi Inoue,\\[-1mm]
\hspace*{2mm} with whom the author has had the great pleasure and privilege of collaborating.\\[-10mm]}
\clearpage

\section{Definitions}

We study simple nondirected $N$-node graphs characterised by adjacency matrices $\bc=\{c_{ij}\}$, with $c_{ij}\!\in\!\{0,1\}$, $c_{ij}\!=\!c_{ji}$ and $c_{ii}\!=\!0$  for all $(i,j)$. They are drawn randomly from maximum entropy  ensembles $p(\bc)$, in which we prescribe the values of all degrees $k_i(\bc)=\sum_j c_{ij}$ and the eigenvalue spectrum $\varrho(\mu|\bc)$, subject to  $\int\!\rmd\mu~\mu\varrho(\mu|\bc)=N^{-1}\sum_{i}c_{ii}=0$ and $\int\!\rmd\mu~\mu^2\varrho(\mu|\bc)=N^{-1}\sum_{ij}c_{ij}c_{ji}=N^{-1}\sum_i k_i$. Since  the spectrum controls the statistics of closed paths of all lengths via its moments, we are not limited to tree-like graphs. Modulo isomorphisms, many graphs are already determined by their spectra \cite{DS1,DS2,Brouwer}, and without external fields the free energy of spin systems on loopy graphs depends only on the loop statistics \cite{Parisi_book}, so we expect that graphs from such ensembles can be tailored very effectively to model complex real-world systems. 
We impose  the degrees as hard constraints, and the spectrum as a soft constraint, so 
\begin{eqnarray}
p(\bc)&\propto & \rme^{N\!\int\!\rmd\mu~\hat{\varrho}(\mu)\varrho(\mu|\bc)}
\prod_{i\leq N}\delta_{k_i,\sum_j c_{ij}}
\label{eq:ensembleA}
\end{eqnarray}
We  can write  the relevant sums over graphs in terms of the generating function
\begin{eqnarray}
\Phi[\hat{\varrho}]&=& \frac{1}{N}\log 
\sum_{\bc}\rme^{N\!\int\!\rmd\mu~\hat{\varrho}(\mu)\varrho(\mu|\bc)}
\prod_{i\leq N}\delta_{k_i,\sum_j c_{ij}}
\label{eq:guzai}
\end{eqnarray}
The equation from which to solve the Lagrange parameter $\hat{\varrho}(\mu)$, i.e. $\varrho(\mu)=\sum_{\bc}p(\bc)\varrho(\mu|\bc)$, 
 and the ensemble entropy per node $S=-N^{-1}\sum_{\bc}p(\bc)\log p(\bc)$, 
 are both expressed in terms of  (\ref{eq:guzai}):
\begin{eqnarray}
\varrho(\mu)=\delta\Phi[\hat{\varrho}]/\delta \hat{\varrho}(\mu),~~~~~~
S=\Phi[\hat{\varrho}]-\int\!\rmd\mu~\hat{\varrho}(\mu)\rho(\mu)
\label{eq:LagrangeSA}
\end{eqnarray}
We could alternatively start by choosing $\hat{\varrho}(\mu)$, and view the first equation of (\ref{eq:LagrangeSA}) as a tool for calculating the associated spectrum. 
Locally tree-like graphs correspond to $\hat{\varrho}(\mu)=0$. For $\hat{\varrho}(\mu)=\alpha_3\mu^3$  with $\alpha_3>0$ we get loopy random graphs constrained by the degrees and the density of triangles. Adding higher order terms to $\hat{\varrho}(\mu)$ means controlling higher order closed path statistics. Imposing the full spectrum means constraining the numbers of closed paths of all lengths. The core of our problem is how to do analytically the sum over graphs in (\ref{eq:guzai}).

\section{Calculation of the generating function}

We can 
 write  (\ref{eq:guzai})  as an average over an Erd\"{os}-R\`{e}nyi graph ensemble \cite{ER} with average degree $\bra k\ket=\frac{1}{N}\sum_i k_i$. Since the probabilities $p_{\rm ER}(\bc)$ of this ensemble depend on $\bc$ only via  $\sum_{i<j}c_{ij}$, 
we can use the short-hand $\bra f(\bc)\ket_{\rm ER}=\sum_{\bc}p_{\rm ER}(\bc)f(\bc)$ to write\begin{eqnarray}
\Phi[\hat{\varrho}]&=& 
  \frac{1}{2}\bra k\ket\big[\log\big(\frac{N}{\bra k\ket}\big)\!+\!1\big]
+
\frac{1}{N}\log \Big\bra
\rme^{N\!\int\!\rmd\mu~\hat{\varrho}(\mu)\varrho(\mu|\bc)}
\prod_{i\leq N}\delta_{k_i,\sum_j c_{ij}}
\Big\ket_{\!\rm ER}\!
+\order(\frac{1}{N})
\label{eq:generator_to_evaluate}
\end{eqnarray}
The factors induced by degree constraints are harmless. Our problem is the dependence  on $\bc$ via $\varrho(\mu|\bc)$. To handle this we use the following identity which can be derived in a few lines from the spectrum formula of \cite{Edwards_Jones_1976}, and was first presented in  \cite{RobertsCoolen}:
\begin{eqnarray}
\rme^{N\!\int\!\rmd\mu~\hat{\varrho}(\mu)\varrho(\mu|\bc)}
&=& 
\lim_{\Delta,\varepsilon\downarrow 0}
\lim_{n_\mu\to \frac{\rmi\Delta}{\pi}\frac{\rmd}{\rmd\mu}\hat{\varrho}(\mu)}
\lim_{m_\mu\to -n_\mu}
\prod_{\mu}\Big[
Z(\mu\!+\!\rmi\varepsilon|\bc)^{n_\mu}~
\overline{Z(\mu\!+\!\rmi\varepsilon|\bc)}^{~m_\mu}
\Big]
\label{eq:main_tool}
\end{eqnarray}
where $\overline{z}$ indicates complex conjugation, and 
with the integrals
$Z(\mu|\bc)=
\int_{\R^N}\!\rmd\bphi~\exp(\!-\!\frac{1}{2}\rmi\bphi\cdot
[\bc\!-\!\mu\one]\bphi)$. 
We can now proceed via the replica method. We evaluate the graph average for integer $\{n_\mu,m_\mu\}$, and  take the limits to imaginary values via analytical continuation\footnote{In standard replica analyses \cite{replicas}  the replica dimension $n$ is taken to zero. Models with nonzero real-valued $n$ describe systems with adiabatically separated timescales and multiple temperatures \cite{Coolen_finite_n1,Coolen_finite_n2,Franz_Mezard,CoolenJort,CoolenUezu,CoolenRabello}. 
Imaginary dimensions $n$ have so far been used only in \cite{Kabashima,Kaba2,Kaba3} and \cite{DerridaMottishaw}.}.
The tricky factor then becomes a product of  integrals:
\begin{eqnarray}
&&
\hspace*{-20mm}
\prod_{\mu}\!\Big[
Z(\mu\!+\!\rmi\varepsilon|\bc)^{n_\mu}~\!
\overline{Z(\mu\!+\!\rmi\varepsilon|\bc)}^{~m_\mu}
\Big]
\nonumber
\\[-3mm]
&=&
\prod_\mu\Big\{
\Big[\!\prod_{\alpha_\mu=1}^{n_\mu}\int_{\R^N}\!\!\rmd\bphi_{\mu,\alpha_\mu}
\rme^{-\frac{1}{2}(\varepsilon-\rmi\mu)\bphi_{\mu,\alpha_\mu}^2}
\Big]
\Big[\!\prod_{\beta_\mu=1}^{m_\mu}\int_{\R^N}\!\!\rmd\bpsi_{\mu,\beta_\mu}
\rme^{-\frac{1}{2}(\varepsilon+\rmi\mu)\bpsi_{\mu,\beta_\mu}^2}
\Big]\Big\}
\nonumber
\\
&&\hspace*{10mm}\times
\rme^{\rmi \sum_{i<j}c_{ij}\sum_\mu\big[
\sum_{\beta_\mu=1}^{m_\mu}\psi_{\mu,\beta_\mu}^i\psi_{\mu,\beta_\mu}^j-
\sum_{\alpha_\mu=1}^{n_\mu}\phi_{\mu,\alpha_\mu}^i\phi_{\mu,\alpha_\mu}^j
\big]}
\label{eq:ZZ}
\end{eqnarray}
We can now do the graph summations. We abbreviate $\bphi^i\!=\!\{\phi_{\mu,\alpha_\mu}^i\}$ and $\bpsi^i=\{\psi_{\mu,\beta_\mu}^i\}$, with
$\bphi^i\!\cdot\bphi^j=\sum_{\mu}\sum_{\alpha_\mu\leq n_\mu}\phi_{\mu,\alpha_\mu}^i\phi_{\mu,\alpha_\mu}^j$ and $\bpsi^i\!\cdot\bpsi^j=\sum_{\mu}\sum_{\beta_\mu\leq m_\mu}\psi_{\mu,\beta_\mu}^i\psi_{\mu,\beta_\mu}^j$, and we introduce a matrix $\bM$ with entries $M_{\mu,\alpha;\mu^\prime,\alpha^\prime}=\mu\delta_{\mu\mu^\prime}\delta_{\alpha\alpha^\prime}$.  Upon inserting (\ref{eq:main_tool}) into (\ref{eq:guzai}), writing degree constraints in integral form via $\delta_{k_i, \sum_j c_{ij}}=(2\pi)^{-1}\int_{-\pi}^\pi \rmd\omega_i~\rme^{\rmi\omega_i(k_i-\sum_j c_{ij})}$, and using (\ref{eq:ZZ}), we find that the order parameter of our problem is the distribution $\Prob(\bphi,\bpsi,\omega)=N^{-1}\sum_i\delta(\bphi\!-\!\bphi^i)\delta(\bpsi\!-\!\bpsi^i)\delta(\omega\!-\!\omega_i)$, which we introduce  into our formulae by inserting for each $(\bphi,\bpsi,\omega)$, with  $\omega\in[-\pi,\pi]$:
\begin{eqnarray}
\hspace*{-0mm}
1&=& \int\!\rmd \Prob(\bphi,\bpsi,\omega)~\delta\Big[\Prob(\bphi,\bpsi,\omega)-\frac{1}{N}\sum_i\delta(\bphi\!-\!\bphi^i)\delta(\bpsi\!-\!\bpsi^i)\delta(\omega\!-\!\omega_i)\Big]
\end{eqnarray}
Upon writing the $\delta$-functions in integral form, 
we then obtain the following path integral representation of $\Phi[\hat{\varrho}]$ which is for large $N$ evaluated via steepest descent, with $\lim_{N\to \infty}\epsilon_N=0$:
\begin{eqnarray}
\Phi[\hat{\varrho}]
&=& \frac{1}{2}\bra k\ket\log\big(\frac{N}{\bra k\ket}\big)
+\lim_{\Delta,\varepsilon\downarrow 0}
\lim_{n_\mu\to \frac{\rmi\Delta}{\pi}\frac{\rmd}{\rmd\mu}\hat{\varrho}(\mu)}
\lim_{m_\mu\to -n_\mu} {\rm extr}_{\{\Prob,\hat{\Prob}\}}\Psi[\Prob,\hat{\Prob}]+\epsilon_N
\label{eq:Phi_steepest_descent}
\\
 \Psi[\Prob,\hat{\Prob}]&=& 
 \rmi\int\!\rmd\bphi\rmd\bpsi\rmd\omega~\hat{\Prob}(\bphi,\bpsi,\omega)\Prob(\bphi,\bpsi,\omega)
   \nonumber
\\
&&\hspace*{-5mm}
+\frac{1}{2}\bra k\ket \!\int\!\rmd\bphi\rmd\bpsi\rmd\omega
\rmd\bphi^\prime\rmd\bpsi^\prime\rmd\omega^\prime ~
\Prob(\bphi,\bpsi,\omega)\Prob(\bphi^\prime\!,\bpsi^\prime\!,\omega^\prime)
\rme^{-\rmi(\omega+\omega^\prime)+\rmi(\bpsi\cdot\bpsi^\prime-\bphi\cdot\bphi^\prime)}
 \nonumber
 \\
  \hspace*{-20mm}
 &&
 \hspace*{-5mm}
 +\sum_k p(k)
\log \int_{-\pi}^\pi\!\frac{\rmd\omega}{2\pi}\rme^{\rmi  k\omega}\! \int\!\!
  \rmd\bphi\rmd\bpsi
~ \rme^{-\frac{1}{2}\bphi\cdot(\varepsilon\one-\rmi\bM)\bphi
  -\frac{1}{2}\bpsi\cdot(\varepsilon\one+\rmi\bM)\bpsi
  -\rmi\hat{\Prob}(\bphi,\bpsi,\omega)}
  \label{eq:Psi_PhatP}
\end{eqnarray}
Working out the saddle-point equations of (\ref{eq:Psi_PhatP})  shows that (\ref{eq:Phi_steepest_descent}) 
  can be written in the form
\begin{eqnarray}
\Phi[\hat{\varrho}]&=& 
 \frac{1}{2}\bra k\ket\big[\log\big(\frac{N}{\bra k\ket}\big) \!+\!1\big]+\sum_k p(k)\log\tilde{p}(k)
+\epsilon_N
\nonumber
\\[-1mm]
&&
+\lim_{\Delta,\varepsilon\downarrow 0}
\lim_{n_\mu\to \frac{\rmi\Delta}{\pi}\frac{\rmd}{\rmd\mu}\hat{\varrho}(\mu)}
\lim_{m_\mu\to -n_\mu}
 \sum_k p(k)
\log  \int\!
  \rmd\bphi\rmd\bpsi~
   \rme^{-\frac{1}{2}\bphi\cdot(\varepsilon\one-\rmi\bM)\bphi
  -\frac{1}{2}\bpsi\cdot(\varepsilon\one+\rmi\bM)\bpsi
 }
 \nonumber
 \\[-1mm]
 &&\hspace*{50mm}\times
  \Big[\int\!\rmd\bphi^\prime \rmd\bpsi^\prime ~
 \WW(\bphi^\prime\!,\bpsi^\prime)\rme^{\rmi(\bpsi\cdot\bpsi^\prime-\bphi\cdot\bphi^\prime)}\Big]^k
  \label{eq:Phi_simplest}
\end{eqnarray}
Here $\tilde{p}(k)=\rme^{-\bra k\ket}\bra k\ket^k/k!$, $\lim_{N\to\infty}\epsilon_N=0$, and $\WW(\bphi,\bpsi)$ is to be solved from 
\begin{eqnarray}
\WW(\bphi,\bpsi)
&=&\sum_{k}\frac{k}{\bra k\ket} p(k) 
\label{eq:closed}
 \\
 &&
\hspace*{-10mm}
 \times
 \frac{
 \rme^{-\frac{1}{2}\bphi\cdot(\varepsilon\one-\rmi\bM)\bphi
  -\frac{1}{2}\bpsi\cdot(\varepsilon\one+\rmi\bM)\bpsi
 }
\Big[\int\!
\rmd\bphi^\prime\rmd\bpsi^\prime~
\WW(\bphi^\prime\!,\bpsi^\prime)
\rme^{\rmi(\bpsi\cdot\bpsi^\prime-\bphi\cdot\bphi^\prime)}
\Big]^{k-1}}
{  \int\!
  \rmd\bphi^\pprime\rmd\bpsi^\pprime
   \rme^{-\frac{1}{2}\bphi^\pprime\cdot(\varepsilon\one-\rmi\bM)\bphi^\pprime
  -\frac{1}{2}\bpsi^\pprime\cdot(\varepsilon\one+\rmi\bM)\bpsi^\pprime
 }
 \Big[\int\!
\rmd\bphi^\prime\rmd\bpsi^\prime~
\WW(\bphi^\prime\!,\bpsi^\prime)
\rme^{\rmi(\bpsi^\pprime\!\cdot\bpsi^\prime-\bphi^\pprime\!\cdot\bphi^\prime)}
\Big]^{k}}
 \nonumber
\end{eqnarray}

\section{Replica symmetric theory} 

\subsection{Replica symmetry ansatz}

We  assume that $\WW(\bphi,\bpsi)$ is symmetric under all permutations of 
$\{\phi_{\mu,1},\ldots,\phi_{\mu,n_\mu}\}$ and  
$\{\psi_{\mu,1},\ldots,\psi_{\mu,m_\mu}\}$. For integer $\{n_\mu,m_\mu\}$ equation (\ref{eq:closed}) is also invariant under 
$\WW(\bphi,\bpsi)\to \overline{\WW(\bpsi,\bphi)}$, and we assume that the relevant solution of  (\ref{eq:closed}) is invariant under this transformation.
It then follows via De Finetti's theorem  \cite{finetti} that
\begin{eqnarray}
\hspace*{-10mm}
\WW(\bphi,\bpsi)&=&\CC \int\! \{\rmd\pi\}
\WW[\{\pi\}]
\Big[\prod_\mu\prod_{\alpha_\mu=1}^{n_\mu}\pi(\phi_{\mu,\alpha_\mu}|\mu)\Big]
\Big[\prod_\mu\prod_{\beta_\mu=1}^{m_\mu}\overline{\pi(\psi_{\mu,\beta_\mu}|\mu)}\Big]
\label{eq:RSansatz_graphs}
\end{eqnarray}
$\CC$ is a constant, and
$\WW[\{\pi\}]$ is a normalised measure on the space of conditioned distributions $\pi$ with $\int\!\rmd x~\pi(x|\mu)=1$ for all $\mu$. We introduce the Fourier transforms $\hat{\pi}(\phi|\mu)=\int\!\rmd x~\rme^{-\rmi x\phi}~\pi(x|\mu)$,  the normalised distributions $ P[\{\pi_1,\ldots,\pi_k\}]$, and the short-hand  $\A[\{\pi_1,\ldots,\pi_k\}]$ as follows:
  \begin{eqnarray}
 P(\phi|\mu,\pi_1,\ldots,\pi_k)&=& \frac{
\rme^{-\frac{1}{2}(\varepsilon-\rmi\mu)\phi^2}
\prod_{\ell\leq k}\hat{\pi}_\ell(\phi|\mu)}
{\int\!\rmd x~
\rme^{-\frac{1}{2}(\varepsilon-\rmi\mu)x^2}
\prod_{\ell\leq k}\hat{\pi}_\ell(x|\mu)
}
\label{eq:define_density_map}
\\
 \A[\{\pi_1,\ldots,\pi_k\}]&=&\prod_\mu\Big[
\Big(\int\!\rmd x~
\rme^{-\frac{1}{2}(\varepsilon-\rmi\mu)x^2}
\prod_{\ell\leq k}\hat{\pi}_\ell(x|\mu)\Big)^{n_\mu}
\Big(\overline{\int\!\rmd x~
\rme^{-\frac{1}{2}(\varepsilon-\rmi\mu)x^2}
\prod_{\ell\leq k}\hat{\pi}_\ell(x|\mu)}\Big)^{m_\mu}
\Big]
\nonumber
\\[-2mm]&&
\label{eq:defineA}
 \end{eqnarray}
This allows us to write equation (\ref{eq:closed}) after some simple manipulations in a form which reveals  that the  ansatz (\ref{eq:RSansatz_graphs}) indeed solves (\ref{eq:closed}),  provided $\WW$ and $\CC$ obey
  \begin{eqnarray}
\WW[\{\pi\}]
&=&\frac{1}{\CC^2}\sum_{k>0}p(k)\frac{k}{\bra k\ket} 
\frac{
\Big[\prod_{\ell<k}\int\!\{\rmd\pi_\ell\}
\WW[\{\pi_\ell\}]
\Big]
 \A[\{\pi_1,\ldots,\pi_{k-1}\}]
\delta_{\rm F}\big[\pi\!-\!P[\{\pi_1,\ldots,\pi_{k-1}\}]\big]
}
{
 \Big[\prod_{\ell\leq k}  \int\!\{\rmd\pi_\ell\}
\WW[\{\pi_\ell\}]
\Big]
 \A[\{\pi_1,\ldots,\pi_k\}]
}
\nonumber
\\[-3mm]&&
\label{eq:Wform}
\\[-2mm]
 \CC^2
&=&\sum_{k>0}p(k)\frac{k}{\bra k\ket}
\frac{
\Big[\prod_{\ell<k}\int\!\{\rmd\pi_\ell\}
\WW[\{\pi_\ell\}]
\Big]
 \A[\{\pi_1,\ldots,\pi_{k-1}\}]
}
{
 \Big[\prod_{\ell\leq k}  \int\!\{\rmd\pi_\ell\}
\WW[\{\pi_\ell\}]
\Big]
 \A[\{\pi_1,\ldots,\pi_k\}]
}
\label{eq:Cform}
\end{eqnarray}
 with the functional delta-distribution $\delta_{\rm F}[g]\propto\prod_x\delta[g(x)]$. 
Our generating function becomes
\begin{eqnarray}
\Phi_{\rm RS}[\hat{\varrho}]&=& 
 \frac{1}{2}\bra k\ket\big[\log\big(\frac{N}{\bra k\ket}\big) \!+\!1\big]+\sum_k p(k)\log\tilde{p}(k)
 +
\bra k\ket \log \CC
 +\epsilon_N
\nonumber
\\
&&
 +\sum_k p(k)
\log
 \Big[\prod_{\ell\leq k} \int\! \{\rmd\pi_\ell\}
\WW[\{\pi_\ell\}]
\Big] \A[\{\pi_1,\ldots,\pi_k\}]
\label{eq:RS_extremum}
\end{eqnarray}
The functional (\ref{eq:defineA}) can be written alternatively in terms of (\ref{eq:define_density_map}) 
 as
\begin{eqnarray}
\A[\{\pi_1,\ldots,\pi_k\}]= \A[P[\{\pi_1,\ldots,\pi_k\}]]
\end{eqnarray}
in which we now define 
\begin{eqnarray}
\A[\{P\}]= \rme^{-\sum_\mu\big[ n_\mu
\log P(0|\mu)
+m_\mu\log\overline{P(0|\mu)}\big]}
\label{eq:AA}
 \end{eqnarray}
This property is due to the fact that $\A[\{\pi_1,\ldots,\pi_k\}]$ can be expressed in terms of the denominators of the right-hand side of (\ref{eq:define_density_map}), which in turn can be written in terms of $P(0|\mu)$ as a consequence of the identity $\hat{\pi}_\ell(0|\mu)=\int\!\rmd x~\pi(x|\mu)=1$ for all $\mu$ and all $\ell$.

\subsection{Formula for the spectrum}

According to (\ref{eq:LagrangeSA}), working out $\varrho(\mu)$ requires differentiation of  $\Phi_{\rm RS}[\hat{\varrho}]$ with respect to $\hat{\varrho}$. This was indeed the origin and purpose of the generating function. This differentiation is easier if we write  $\Phi_{\rm RS}[\hat{\varrho}]$ as an extremum over $\CC$ and $\WW$, so that the dependences of $\CC$ and $\WW$ on $\hat{\varrho}$ will not affect these derivatives. If we also extremise over $\CC$ for fixed $\WW$, and eliminate $\CC$, we obtain
\begin{eqnarray}
 \CC^{-2}
 &=&
 \int\! \{\rmd\pi\rmd\pi^\prime\}
\WW[\{\pi\}]\WW[\{\pi^\prime\}] \BB[\{\pi,\pi^\prime\}]
\label{eq:Csolved}
\\
 \Phi_{\rm RS}[\hat{\varrho}]&=&
 \frac{1}{2}\bra k\ket\big[\log\big(\frac{N}{\bra k\ket}\big)\!+\!1\big]+\sum_k p(k)\log\tilde{p}(k)+\epsilon_N
\nonumber
\\
&&\hspace*{-0mm}
+~
{\rm extr}_{\{\WW\}}\Big\{
\sum_k p(k)
\log\Big[
 \Big[\prod_{\ell\leq k}
 \int\! \{\rmd\pi_\ell\}
\WW[\{\pi_\ell\}]\Big]
 \A[\{\pi_1,\ldots,\pi_k\}]
\Big]
\nonumber
\\
&&\hspace*{10mm}
-\frac{1}{2}\bra k\ket \log  \int\! \{\rmd\pi\rmd\pi^\prime\}
\WW[\{\pi\}]\WW[\{\pi^\prime\}] \BB[\{\pi,\pi^\prime\}]
\Big\}
\label{eq:intermediate3_entropy_direct}
\end{eqnarray}
\vspace*{-6mm}

\noindent
with 
\begin{eqnarray}
\hspace*{-5mm}
\BB[\{\pi,\pi^\prime\}]&=& 
\prod_\mu \Big\{
\Big[ \int\!\rmd \phi\rmd \phi^\prime\pi(\phi|\mu)\pi^\prime(\phi^\prime|\mu)\rme^{-\rmi \phi \phi^\prime}\Big]^{n_\mu}
\Big[
\overline{\int\!\rmd \phi\rmd \phi^\prime\pi(\phi|\mu)\pi^\prime(\phi^\prime|\mu)\rme^{-\rmi \phi \phi^\prime}}
\Big]^{m_\mu}
\label{eq:BB}
\end{eqnarray}
Now $\hat{\varrho}$ appears only in $\A[\ldots]$ and $\BB[\ldots]$, and we obtain from (\ref{eq:LagrangeSA}) the following formula:
\begin{eqnarray}
\varrho_{\rm RS}(\mu)&=& 
\sum_k p(k)\int\!\{\rmd\pi\rmd\pi^\prime\}\WW[\{\pi^\prime\}]\BB[\{\pi,\pi^\prime\}] \Big\{
\Big( \frac{\delta\log\A[\{\pi\}] }{\delta\hat{\varrho}(\mu)}
+
\frac{\delta\log \BB\big[\{\pi,\pi^\prime\}] }{\delta\hat{\varrho}(\mu)}\Big)
\nonumber
\\
&&\hspace*{15mm}
\times\frac{
 \int\! 
 \Big[\prod_{\ell< k}
\{\rmd\pi_\ell\}
\WW[\{\pi_\ell\}]\Big] \delta_{\rm F}\big[\pi-P[\{\pi_1,\ldots,\pi_{k-1}\}]\big]
\A[\{\pi\}] 
}{
\int\!
 \Big[\prod_{\ell\leq k}
 \int\! \{\rmd\pi_\ell\}
\WW[\{\pi_\ell\}]\Big]
\A[\{\pi_1,\ldots,\pi_k\}]
}
\Big\}
\nonumber
\\
&&\hspace*{5mm}
-\frac{1}{2}\bra k\ket \frac{ \int\! \{\rmd\pi\rmd\pi^\prime\}
\WW[\{\pi\}]\WW[\{\pi^\prime\}]\BB[\{\pi,\pi^\prime\}] \delta\log \BB[\{\pi,\pi^\prime\}]/\delta\hat{\varrho}(\mu)}
{ \int\! \{\rmd\pi\rmd\pi^\prime\}
\WW[\{\pi\}]\WW[\{\pi^\prime\}] \BB[\{\pi,\pi^\prime\}]}
\label{eq:spectrum_in_pi_1}
\end{eqnarray}
Substituting the definitions of $n_\mu$ and $m_\mu$, and taking $\Delta\downarrow 0$,  
the functional derivatives become
\begin{eqnarray}
 \frac{\delta\log\A[\{\pi\}] }{\delta\hat{\varrho}(\mu)}&=& 
 -\frac{2}{\pi}\frac{\rmd}{\rmd\mu} {\rm Arg}~ \pi(0|\mu)
 \label{eq:logA_general}
 \\
 \frac{\delta\log \BB[\{\pi,\pi^\prime\}]}{\delta\hat{\varrho}(\mu)}&=&
\frac{2}{\pi} \frac{\rmd}{\rmd\mu}{\rm Arg}\int\!\rmd \phi\rmd \phi^\prime \rme^{-\rmi \phi \phi^\prime}
\pi(\phi|\mu)\pi^\prime(\phi^\prime|\mu)
\label{eq:logB_general}
\end{eqnarray}

 \section{Exploiting the nature of the propagation}

\subsection{Gaussian propagated densities}

Equation (\ref{eq:Wform}) can be seen as the fixed-point equation of a process in which densities $\pi(x|\mu)$ are mapped via (\ref{eq:define_density_map}). As in e.g. \cite{Kuehn}, this dynamics is seen to close for complex Gaussian functions:
\begin{eqnarray}
\WW[\{\pi\}]&=& \int\{\rmd x\rmd y\}\WW[\{x,y\}]~\delta_{\rm F}\big[\pi-\pi[\{x,y\}]\big]
\label{eq:gaussian_ansatz}
\end{eqnarray}
Here $\{x\}$ and $\{y\}$ are complex functions of $\mu\in\R$, with ${\rm Im}~x(\mu)<0$ for all $\mu$,  and $\pi[\{x,y\}]$ is the $\mu$-conditioned normalised density $\pi(\phi|x(\mu),y(\mu))$ defined via  
\begin{eqnarray}
\pi(\phi|x,y)=\frac{\rme^{-\frac{1}{2}\rmi x\phi^2+\rmi y\phi}}{Z(x,y)},
\label{eq:pi_gaussian}
~~~~~~
Z(x,y)=\Big(\frac{2\pi}{|x|}\Big)^{\!\frac{1}{2}}\rme^{\frac{1}{2}\rmi ~\atan[{\rm Re}(x)/{\rm Im}(x)]+\frac{1}{2}\rmi y^2/x}
\end{eqnarray}
The integrals leading to the normalisation factor $Z(x,y)$ are found in \cite{GR}. 
The Fourier transform of (\ref{eq:pi_gaussian}) is
$\hat{\pi}(\phi|\mu)=\exp[\frac{1}{2}\rmi \phi^2/x(\mu)- \rmi \phi y(\mu)/x(\mu)]$, so (\ref{eq:define_density_map}) takes the form
\begin{eqnarray}
 P(\phi|\mu,\pi_1,\ldots,\pi_{k-1})&=& \pi(\phi|F(\mu|x_1,\ldots,x_{k-1}),G(\mu|x_1,y_1,\ldots,x_{k-1},y_{k-1}))
\end{eqnarray}
Here $F(\ldots)$ and $G(\ldots)$ are the following functions of $\mu$:
\begin{eqnarray}
F(\mu|x_1,\ldots, x_{k-1})&=& -\mu-\rmi \varepsilon-
\sum_{\ell<k}1/x_\ell(\mu)
\label{eq:Fmap}
\\
G(\mu|x_1,y_1,\ldots, x_{k-1},y_{k-1})&=&  - \sum_{\ell<k}y_\ell(\mu)/x_\ell(\mu)
\label{eq:Gmap}
\end{eqnarray}
If all $x_\ell(\mu)$ have negative imaginary parts, so will $F(\mu|x_1,\ldots,x_{k-1})$. Hence all relevant integrals over $\phi$ exist and the propagation  (\ref{eq:define_density_map}) indeed closes within the family (\ref{eq:pi_gaussian}). 

\subsection{Order parameter equations and spectrum formula}

We work out the functionals $\A[\{\pi\}]$ and $\BB[\{\pi,\pi^\prime\}]$. Starting from (\ref{eq:defineA},\ref{eq:BB}) or 
(\ref{eq:logA_general},\ref{eq:logB_general}) one derives
\begin{eqnarray}
\A[\{x,y\}]&=& \rme^{
 \frac{1}{\pi}\int\!\rmd\mu~\hat{\varrho}(\mu)\frac{\rmd}{\rmd\mu} 
 \big[\atan\big(\frac{{\rm Re}(x(\mu))}{{\rm Im}(x(\mu))}\big)+{\rm Re}\big(\frac{y^2(\mu)}{x(\mu)}\big)\big]}
\label{eq:Afinal_gauss}
 \\[1mm]
  \BB[\{x,y;x^\prime,y^\prime\}]&=&
 \rme^{\frac{1}{\pi}
 \int\!\rmd\mu~\hat{\varrho}(\mu)
 \frac{\rmd}{\rmd\mu}\big[
 \atan\big(\frac{{\rm Re}(x(\mu)-1/x^\prime(\mu))}{{\rm Im}(x(\mu)-1/x^\prime(\mu))}\big)
 -
 \atan\big(\frac{{\rm Re}(x(\mu))}{{\rm Im}(x(\mu))}\big)
 \big]}
 \nonumber
 \\
 &&\times 
  \rme^{\frac{1}{\pi}
 \int\!\rmd\mu~\hat{\varrho}(\mu) \frac{\rmd}{\rmd\mu}
 {\rm Re}\big(
\frac{y^2(\mu)/x(\mu)+y^{\prime 2}(\mu)/x^\prime(\mu)-2y(\mu)y^\prime(\mu)}{x(\mu)x^\prime(\mu)-1}
\big) 
 }
\label{eq:Bfinal_gauss}
\end{eqnarray}
Similarly we can work out the order parameter equations (\ref{eq:Wform}) for $\WW[\{x,y\}]$ and (\ref{eq:Csolved}) for the normalisation factor $\CC$. To compactify the result we introduce the short-hand 
\begin{eqnarray}
\FF_k[\{x,y\}]&=& \Big[\prod_{\ell\leq k}\int\!\{\rmd x_\ell \rmd y_\ell\}
\WW[\{x_\ell,y_\ell\}]
\Big]
\delta_{\rm F}\Big[
\!\!
\begin{array}{l}
x\!-\!F[x_1,\ldots,x_{k}]
\\
y\!-\!G[x_1,y_1,\ldots,x_{k},y_{k}]
\end{array}
\!\!
\Big]
\label{eq:F_K0}
\end{eqnarray}
Here $F[x_1,\ldots,x_k]$ and $G[x_1,y_1,\ldots,x_{k},y_{k}]$ are the functions defined in (\ref{eq:Fmap},\ref{eq:Gmap}). 
Now
  \begin{eqnarray}
\WW[\{x,y\}]
&=&\frac{1}{\CC^2}\sum_{k>0}p(k)\frac{k}{\bra k\ket} 
\frac{\A[\{x,y\}]\FF_{k-1}[\{x,y\}]}
{
\int\!\{\rmd x^\prime\rmd y^\prime\}
\A[\{x^\prime,y^\prime\}]
\FF_k[\{x^\prime,y^\prime\}]
}
\label{eq:SPE_W_gaussian}
\\
 \CC^{-2}
 &=&
 \int\! \{\rmd x\rmd y\rmd x^\prime \rmd y^\prime\}
\WW[\{x,y\}]\WW[\{x^\prime,y^\prime\}] \BB[\{x,y;x^\prime,y^\prime\}]
\label{eq:SPE_C_gaussian}
\end{eqnarray}
Finally we turn to our spectrum equation (\ref{eq:spectrum_in_pi_1}), which for the density (\ref{eq:gaussian_ansatz}) becomes
\begin{eqnarray}
\varrho_{\rm RS}(\mu)&=& 
\sum_k p(k)\int\!\{\rmd x\rmd y\rmd x^\prime\rmd y^\prime\}\WW[\{x^\prime,y^\prime\}]\BB[\{x,y;x^\prime,y^\prime\}] 
\nonumber
\\
&&\hspace*{-0mm}
\times
\frac{A[\{x,y\}]\FF_{k-1}[\{x,y\}]
}{
\int\!\{\rmd x^\prime\rmd y^\prime\}
A[\{x^\prime,y^\prime\}]
\FF_k[\{x^\prime,y^\prime\}]}
\Big( \frac{\delta\log\A[\{x,y\}] }{\delta\hat{\varrho}(\mu)}
+
\frac{\delta\log \BB\big[\{x,y;x^\prime,y^\prime\}] }{\delta\hat{\varrho}(\mu)}\Big)
\nonumber
\\[2mm]
&&\hspace*{-0mm}
-\frac{1}{2}\bra k\ket \frac{ \int\! \{\rmd x\rmd y\rmd x^\prime\rmd y^\prime\}
\WW[\{x,y\}]\WW[\{x^\prime,y^\prime\}]~\delta\BB[\{x,y;x^\prime,y^\prime\}]/\delta\hat{\varrho}(\mu)}
{ \int\! \{\rmd x\rmd y\rmd x^\prime\rmd y^\prime\}
\WW[\{x,y\}]\WW[\{x^\prime,y^\prime\}] \BB[\{x,y;x^\prime,y^\prime\}]}
\label{eq:spectrum_in_xy}
\end{eqnarray}
with\\[-10mm]
\begin{eqnarray}
\frac{\delta\log\A[\{x,y\}]}{\delta\hat{\varrho}(\mu)}&=& 
 \frac{1}{\pi}\frac{\rmd}{\rmd\mu} 
 \Big[\atan\big(\frac{{\rm Re}(x)}{{\rm Im}(x)}\big)+{\rm Re}\big(\frac{y^2(\mu)}{x(\mu)}\big)\Big]
\\[1mm]
\frac{\delta\log\BB[\{x,y;x^\prime,y^\prime\}]}{\delta\hat{\varrho}(\mu)}&=& 
\frac{1}{\pi}
 \frac{\rmd}{\rmd\mu}\Big[
 \atan\big(\frac{{\rm Re}(x(\mu)-1/x^\prime(\mu))}{{\rm Im}(x(\mu)-1/x^\prime(\mu))}\big)
 -
 \atan\big(\frac{{\rm Re}(x(\mu))}{{\rm Im}(x(\mu))}\big)
 \nonumber
 \\
 &&
 +
 {\rm Re}\big(
\frac{y^2(\mu)/x(\mu)+y^{\prime 2}(\mu)/x^\prime(\mu)\!-\!2y(\mu)y^\prime(\mu)}{x(\mu)x^\prime(\mu)-1}
\big) \Big]
\end{eqnarray}
  
\subsection{Reflection symmetry in the origin}

Our equations are invariant under 
$\WW[\{x,y\}]\to \WW[\{x,-y\}]$, which represents reflection in the origin of the complex plane of the centres of the $\pi(\phi|x,y)$. 
Assuming the true saddle-point to be of the invariant form $\WW[\{x,y\}]=\WW[\{x\}]\delta[\{y\}]$ implies that we consider the functions  $\pi(\phi|x,y)$ to be centred at the origin, and we will find that this is a property that holds in the limit of treelike ensembles. 
It simplifies our equations to
  \begin{eqnarray}
\WW[\{x\}]&=& \frac{1}{\CC^2}
\sum_{k>0}p(k)\frac{k}{\bra k\ket}
 \frac{ \A[\{x\}]\FF_{k-1}[\{x\}]
}
{
\int\!\{\rmd x^\prime\}
\A[\{x^\prime\}]\FF_k[\{x^\prime\}]}
\\
\hspace*{7mm} \CC^{-2}
&=& 
\int\{\rmd x\rmd x^\prime\}~\WW[\{x\}]\WW[\{x^\prime\}]\BB[\{x,x^\prime\}]
\end{eqnarray} 
with $\FF_k[\{x,y\}]=\FF_k[\{x\}]\delta[\{y\}]$, and 
\begin{eqnarray}
\A[\{x\}]&=& \rme^{
 \frac{1}{\pi}\int\!\rmd\mu~\hat{\varrho}(\mu)\frac{\rmd}{\rmd\mu} 
 \atan\big(\frac{{\rm Re}(x(\mu))}{{\rm Im}(x(\mu))}\big)}
\label{eq:Afinal_gauss_x}
 \\[1mm]
  \BB[\{x;x^\prime\}]&=&
 \rme^{\frac{1}{\pi}
 \int\!\rmd\mu~\hat{\varrho}(\mu)
 \frac{\rmd}{\rmd\mu}\big[
 \atan\big(\frac{{\rm Re}(x(\mu)-1/x^\prime(\mu))}{{\rm Im}(x(\mu)-1/x^\prime(\mu))}\big)
 -
 \atan\big(\frac{{\rm Re}(x(\mu))}{{\rm Im}(x(\mu))}\big)
 \Big]}
\label{eq:Bfinal_gauss_x}
\end{eqnarray}
The spectrum becomes
\begin{eqnarray}
\varrho_{\rm RS}(\mu)&=& 
\sum_k p(k)\int\!\{\rmd x\rmd x^\prime\}\WW[\{x^\prime\}]\BB[\{x;x^\prime\}] 
\nonumber
\\[-2mm]
&&\hspace*{25mm}
\times \frac{A[\{x\}]\FF_{k-1}[\{x\}]
}{
\int\!\{\rmd x^\prime\}
A[\{x^\prime\}]\FF_k[\{x^\prime\}]
}
\Big( \frac{\delta\log\A[\{x\}] }{\delta\hat{\varrho}(\mu)}
+
\frac{\delta\log \BB\big[\{x;x^\prime\}] }{\delta\hat{\varrho}(\mu)}\Big)
\nonumber
\\[2mm]
&&\hspace*{0mm}
-\frac{1}{2}\bra k\ket \frac{ \int\! \{\rmd x\rmd x^\prime\}
\WW[\{x\}]\WW[\{x^\prime\}] ~\delta \BB[\{x;x^\prime\}]/\delta\hat{\varrho}(\mu)}
{ \int\! \{\rmd x\rmd x^\prime\}
\WW[\{x\}]\WW[\{x^\prime\}] \BB[\{x;x^\prime\}]}
\label{eq:spectrum_in_x}
\end{eqnarray}
Continuous 
bifurcations of states with $\WW[\{x,y\}]\neq \WW[\{x\}]\delta[\{y\}]$ can be located via a Guzai (i.e. functional moment) expansion \cite{guzaipaper}. One can show that such bifurcations do occur, but it is not yet clear whether they correspond to physically relevant transitions. 

\subsection{Regular graphs}

For regular graphs our formulae simplify considerably. Here we find the order parameter equation (\ref{eq:SPE_W_gaussian}) for $\WW[\{x,y\}]$ and the spectrum reducing to
  \begin{eqnarray}
\WW[\{x,y\}]
&=&\frac{
\A[\{x,y\}]\FF_{k-1}[\{x,y\}]
}
{
\int\!\{\rmd x^\prime\rmd y^\prime\}
\A[\{x^\prime,y^\prime\}]
\FF_{k-1}[\{x^\prime,y^\prime\}]}
\label{eq:SPE_W_xy_regular}
\\[1mm]
\varrho_{\rm RS}(\mu)&=& 
\frac{\int\!\{\rmd x\rmd y\rmd x^\prime\rmd y^\prime\}\WW[\{x,y\}]\WW[\{x^\prime,y^\prime\}]\BB[\{x,y;x^\prime,y^\prime\}] ~
\varrho(\mu|\{x,y;x^\prime,y^\prime\})
}
 {\int\!\{\rmd x\rmd y\rmd x^\prime\rmd y^\prime\}\WW[\{x,y\}]\WW[\{x^\prime,y^\prime\}]\BB[\{x,y;x^\prime,y^\prime\}] }
\label{eq:spectrum_in_xy_regular}
\end{eqnarray}
with 
\begin{eqnarray}
\varrho(\mu|\{x,y;x^\prime,y^\prime\})&=& 
  \frac{1}{2\pi}\frac{\rmd}{\rmd\mu}\Big\{
  (1\!-\!\frac{1}{2}k)\Big[
 \atan\big(\frac{{\rm Re}(x(\mu)\!-\!1/x^\prime(\mu))}{{\rm Im}(x(\mu)\!-\!1/x^\prime(\mu))}\big)
  \!+\! 
 \atan\big(\frac{{\rm Re}(x^\prime(\mu)\! -\! 1/x(\mu))}{{\rm Im}(x^\prime(\mu)\! -\! 1/x(\mu))}\big)
 \Big]
 \nonumber
 \\
&&\hspace*{10mm}
+\frac{1}{2}k\Big[ \atan\big(\frac{{\rm Re}(x(\mu))}{{\rm Im}(x(\mu))}\big)
+\atan\big(\frac{{\rm Re}(x^\prime(\mu))}{{\rm Im}(x^\prime(\mu))}\big)
\Big]
\nonumber
\\
&&\hspace*{-15mm}
 +
 {\rm Re}\Big(
(2\!-\!k)\frac{y^2(\mu)/x(\mu)\!+\!y^{\prime 2}(\mu)/x^\prime(\mu)\!-\!2y(\mu)y^\prime(\mu)}{x(\mu)x^\prime(\mu)-1}
\!+\!\frac{y^2(\mu)}{x(\mu)}\!+\!\frac{y^{\prime 2}(\mu)}{x^\prime(\mu)}
\Big)
 \Big\}
 \label{eq:spectrum_regular_inside}
\end{eqnarray}

\section{Interpretation and simple tests of the theory}

\subsection{Interpretation and link with message passing algorithms}

In the  limit $\hat{\varrho}\to 0$, where $\CC=\A[\{x,y\}]=1$ for all $\{x,y\}$,  formula
  (\ref{eq:SPE_W_gaussian})  acquires  the standard form $\WW[\{x,y\}]=\sum_{k>0}p(k)(k/\bra k\ket)\FF_{k-1}[\{x,y\}]$ of message passing algorithms on tree-like graphs. 
Also the more complex structure of   (\ref{eq:SPE_W_gaussian})  can be interpreted as describing the stationary state of a message passing process, but now extended with the nontrivial message acceptance probabilities
   \begin{eqnarray}
  \Prob[\{x,y\}|k]&\propto&
 \frac{ \A[\{x,y\}]
}
{
\int\!\{\rmd x^\prime\rmd y^\prime\}
\A[\{x^\prime\!,y^\prime\}]\FF_k[\{x^\prime\!,y^\prime\}]
}
\label{eq:accept}
\end{eqnarray}
(modulo a multiplicative constant). This is similar to the re-weighting of solutions in e.g. \cite{SAT}. 
In  tree-like graphs we accept {\em each} proposed message $\{x,y\}$. 
This interpretation shows us how to solve (\ref{eq:SPE_W_gaussian},\ref{eq:SPE_C_gaussian}) numerically, and  how we should adapt belief propagation and other message passing algorithms \cite{belief1,belief2,montanari},  to make these exact for stochastic processes on loopy graphs.  

   \subsection{Locally tree-like graphs}
   
Our theory 
 should recover established results on locally tree-like graphs upon setting $\hat{\varrho}=0$. This reduces our ensemble to a maximum entropy one in which only the degrees are prescribed. Now $\A[\{x,y\}]\!=\!\BB[\{x,y;x^\prime\!,y^\prime\}]\!=\!\CC\!=\!1$, and 
we indeed recover from (\ref{eq:RS_extremum}) 
the correct  ensemble entropy density of tree-like ensembles with prescribed degrees  \cite{Annibale_etal},   with $\lim_{N\to\infty}\epsilon_N=0$:
 \begin{eqnarray}
S&=&
    \frac{1}{2}\bra k\ket\big[\log\big(\frac{N}{\bra k\ket}\big)\!+\!1\big]+\sum_k p(k)\log\tilde{p}(k)     +\epsilon_N
    \end{eqnarray}
  The spectra of locally tree-like graphs are also recovered correctly, with solutions of the simple form $\WW[\{x,y\}]=\WW[\{x\}]\delta[\{y\}]$ and real-valued $\{x\}$. Our present formalism 
  predicts that  \begin{eqnarray}
 \varrho(\mu)&=&- \frac{\rmd}{\rmd\mu}\Big\{
\frac{1}{2}\sum_k p(k)
\int\!\{ \rmd x\}\FF_k[\{x\}]
{\rm sgn}\big[x(\mu)\big]
\Big]
\label{eq:simple_tree_spectrum}
\\
&&
 +\frac{1}{2}\bra k\ket 
 \int\! \{
 \rmd x\rmd x^\prime
 \}
\WW[\{x\}]\WW[\{x^\prime\}] ~
\theta[x(\mu)x^\prime(\mu)]\theta[1\!-\!x(\mu)x^\prime(\mu)]{\rm sgn}[x(\mu)\!+\!x^\prime(\mu)]
  \Big\}
  \nonumber
  \\[-6.5mm]&&\nonumber
\end{eqnarray}
in which
\begin{eqnarray}
\WW[\{x\}]= 
\sum_{k>0}p(k)\frac{k}{\bra k\ket} 
  \FF_{k-1}[\{x\}],~~~~~~
\FF_k[\{x\}]= \Big[\prod_{\ell\leq k}\int\!\{\rmd x_\ell\}\WW[\{x_\ell\}]\Big]\delta_{\rm F}\big[
x\!-\!F[x_1,\ldots,x_k]
\big]
  \label{eq:simple_tree_orderparameter}
\end{eqnarray} 
 For $k$-regular tree-like graphs our order parameter equation (\ref{eq:simple_tree_orderparameter}) 
 and the spectrum  (\ref{eq:simple_tree_spectrum}) become 
\begin{eqnarray}
\WW[\{x\}]&=& 
 \Big[\prod_{\ell< k}\int\!\{\rmd x_\ell\}\WW[\{x_\ell\}]\Big]\delta_{\rm F}\big[
x\!-\!F[x_1,\ldots,x_{k-1}]
\big]
\label{eq:orderpar_treelike_regular}
\\
 \varrho(\mu)&=&
 - \frac{1}{2}\frac{\rmd}{\rmd\mu}\int\!\{\rmd x\rmd x^\prime\}\WW[\{x\}]\WW[\{x^\prime\}]~
\nonumber
\\
&&\times  \theta[x(\mu)x^\prime(\mu)]{\rm sgn}[x(\mu)\!+\!x^\prime(\mu)]
 \Big[1\!+\!
  (k\!-\!2)
\theta[1\!-\!x(\mu)x^\prime(\mu)]  
\Big]
\label{eq:spectrum_treelike_regular}
\end{eqnarray}
For $k\!=\!1$ equation (\ref{eq:orderpar_treelike_regular}) immediately leads to $ \varrho(\mu)=
  \frac{1}{2}\delta(\mu\!-\!1)+ \frac{1}{2}\delta(\mu\!+\!1)$, 
which is indeed the spectrum of a graph consisting of $N/2$ disconnected 2-node components (the only possible regular graph with $k=1$). Equation (\ref{eq:orderpar_treelike_regular}) can be solved analytically also for $k>1$. To find  the spectrum we only need the distribution $\WW(x|\mu)$ of each individual $x(\mu)$, 
for which we find:
\begin{eqnarray}
\hspace*{-5mm}
|\mu|<2\sqrt{k\!-\!1}:&~~~&
\WW(x|\mu)=
\frac{1}{\pi}
\frac{\sqrt{k\!-\!1\!-\!\frac{1}{4}\mu^2}}{(x\!+\!\frac{1}{2}\mu)^2+k\!-\!1\!-\!\frac{1}{4}\mu^2}
\label{eq:Wsoln_treelike_regular1}
\\
\hspace*{-5mm}
|\mu|>2\sqrt{k\!-\!1}:&~~~&
\WW(x|\mu)=\delta\Big[x\!+\!\frac{1}{2}\mu\!+\!\frac{1}{2}\mu\sqrt{1\!-\!4(k\!-\!1)/\mu^2}\Big]
\label{eq:Wsoln_treelike_regular2}
\end{eqnarray}
Inserting this result into (\ref{eq:spectrum_treelike_regular}) leads us to an integral that can again be evaluated analytically, and  recovers 
the correct spectrum of \cite{McKay} for $k$-regular treelike graphs:
 \begin{eqnarray}
\varrho(\mu)&=& \theta\big[2\sqrt{k\!-\!1}\!-\!|\mu|\big]~\frac{k\sqrt{4(k\!-\!1)\!-\!\mu^2}}{2\pi(k^2\!-\!\mu^2)}
\label{eq:McKay}
\end{eqnarray}
For graphs with arbitrary degree distributions $p(k)$ we can show  that from   (\ref{eq:simple_tree_orderparameter}) indeed follows the equation obtained in \cite{Dorogovtsev}. Thus our formalism passes the available tests in the tree-like limit.

\section{Current and next stage of the research programme}

\subsection{Equations for order parameters and spectra of loopy ensembles}

We now also have explicit equations for the analysis of graphs with extensively many short loops, where $\hat{\varrho}(\mu)\neq 0$. 
The order parameter equation and the spectrum are given by (\ref{eq:SPE_W_xy_regular},\ref{eq:spectrum_in_xy_regular}). 
In the tree-like limit we had $\WW[\{x,y\}]=\WW[\{x\}]\delta[\{y\}]$ with ${\rm Im}[x(\mu)]\uparrow 0$ for all $\mu$. 
If we make the ansatz that also in the presence of loops we still have 
$\WW[\{x,y\}]=\WW[\{x\}]\delta[\{y\}]$ with ${\rm Im}[x(\mu)]\uparrow 0$ for all $\mu$, and limit ourselves to regular graphs, we can simplify our order parameter equation to
 \begin{eqnarray}
\WW[\{x\}]
&=&\frac{
\A[\{x\}]\Big[\prod_{\ell< k}\int\!\{\rmd x_\ell \}
\WW[\{x_\ell\}]
\Big]
\delta_{\rm F}\big[
x\!-\!F[x_1,\ldots,x_{k-1}]
\big]
}
{
\int\!\{\rmd x^\prime\}
\A[\{x^\prime\}]
\Big[\prod_{\ell< k}\int\!\{\rmd x_\ell \}
\WW[\{x_\ell\}]
\Big]
\delta_{\rm F}\big[
x^\prime\!-\!F[x_1,\ldots,x_{k-1}]
\big]
}
\label{eq:SPE_W_xy_regular_repeat}
\end{eqnarray}
with $\A[\{x\}]=\exp\big[\!-\!\frac{1}{2}\int\!\rmd\mu~\hat{\varrho}(\mu)\frac{\rmd}{\rmd\mu} 
 {\rm sgn}[x(\mu)]\big]$. 
The function $F[x_1,\ldots,x_{k-1}]$ is for $\varepsilon\downarrow 0$ defined as
$F(\mu|x_1,\ldots, x_{k-1})= -\mu-
\sum_{\ell<k}1/x_\ell(\mu)$. 
The spectrum formula simplifies to
\begin{eqnarray}
\varrho_{\rm RS}(\mu)&=& 
\frac{\int\!\{\rmd x\rmd x^\prime\}\WW[\{x\}]\WW[\{x^\prime\}]\BB[\{x;x^\prime\}] ~
\varrho(\mu|\{x;x^\prime\})
}
 {\int\!\{\rmd x\rmd x^\prime\}\WW[\{x\}]\WW[\{x^\prime\}]\BB[\{x;x^\prime\}] }
\label{eq:spectrum_in_xy_regular_repeat}
\\
\varrho(\mu|\{x;x^\prime\})&=& 
 - \frac{1}{4}\frac{\rmd}{\rmd\mu}\Big\{
\Big[ {\rm sgn}[x(\mu)]  \!+\!   {\rm sgn}[x^\prime(\mu)] \Big]
\Big[  
 1\!+\!(k\!-\!2)
 \theta[1\!-\!x(\mu)x^\prime(\mu)]
\Big]
 \Big\}
 \label{eq:spectrum_regular_inside_repeat}
\end{eqnarray}
in which
\begin{eqnarray}
  \BB[\{x;x^\prime\}]&=&
 \rme^{
\frac{1}{2} \int\!\rmd\mu~\hat{\varrho}(\mu)
 \frac{\rmd}{\rmd\mu}\big(
 \big[{\rm sgn}[x(\mu)]+{\rm sgn}[x^\prime(\mu)]\big]
\theta [1-x(\mu)x^\prime(\mu)]
\big)
 }
\label{eq:Bfinal_gauss_repeat}
\end{eqnarray}
It is not difficult to prove that (\ref{eq:spectrum_regular_inside_repeat}) is normalised  correctly. 
However, testing the validity of our equations in full is nontrivial for a number of reasons. 

First, there appear to exist no benchmark solutions yet, in the form of exact and independent asymptotic spectrum derivations for ensembles of the form (\ref{eq:ensembleA}) with $\hat{\varrho}\neq 0$, against which to test our predictions. Second, solving the order parameter equations numerically for $\hat{\varrho}\neq 0$ is nontrivial, as it involves population dynamics algorithms in which we propagate functions rather than fields, and with nontrivial acceptance probabilities, that are quite hard to equilibrate accurately without sacrificing eigenvalue resolution. Thirdly, generating random graphs from (\ref{eq:ensembleA}) numerically, in order to then obtain their spectra by numerical diagonalisation,  is itself nontrivial \cite{CoolenMartinoAnnibale}; 
for relatively simple choices like $\hat{\varrho}(\mu)=\alpha_3 \mu^3\!+\!\alpha_4 \mu^4$ reliable algorithms and code do exist, but these ensembles are notorious for their complex phase transitions \cite{Strauss,Triangles,Burda1,Burda2}. For instance, for regular graphs with $k=3$ and 
$\hat{\varrho}(\mu)=\alpha_3 \mu^3$ one observes in numerical simulations  that, modulo finite size effects,
the spectrum is of the form
\begin{eqnarray}
\varrho(\mu)&\approx& (1\!-\gamma) \tilde{\varrho}(\mu)+\gamma\Big\{
\frac{3}{4}\delta(\mu\!+\!1)+\frac{1}{4}\delta(\mu\!-\!3)\Big\}
\label{eq:observed_a3}
\end{eqnarray}
in which $\gamma\in[0,1]$ and $\tilde{\varrho}(\mu)$ is McKay's $k=3$ formula (\ref{eq:McKay}) for tree-like regular random graphs. One confirms that $\bra \mu\ket=0$ and $\bra \mu^2\ket=3$ for any $\gamma$, as required. The second term in (\ref{eq:observed_a3}) is the spectrum of a  fully connected simple graphlet of size four. Hence  (\ref{eq:observed_a3})  describes graphical phase separation; the system  increases the number of triangles by disconnecting a fraction $\gamma$ of the nodes from the treelike bulk, to form $\gamma N/4$ disconnected 4-node cliques. 
Similarly, For regular graphs with $k=3$ and 
$\hat{\varrho}(\mu)=\alpha_4 \mu^4$ one observes in numerical simulations  spectra of the form
\begin{eqnarray}
\varrho(\mu)&\approx& (1\!-\gamma) \tilde{\varrho}(\mu)+\gamma\Big\{
\frac{2}{3}\delta(\mu)+\frac{1}{6}\delta(\mu\!-\!3)+\frac{1}{6}\delta(\mu\!-\!3)\Big\}
\label{eq:observed_a4}
\end{eqnarray}
in which again $\tilde{\varrho}(\mu)$ is McKay's  formula (\ref{eq:McKay}) for $k=3$. Also here $\bra \mu\ket=0$ and $\bra \mu^2\ket=3$ for any $\gamma$. The second term of (\ref{eq:observed_a4}) is  the spectrum of the six node graphlet with adjacency matrix $a_{ij}=\delta_{i,j+1}+\delta_{i,j-1}+\delta_{i,j+3}+\delta_{i,j-3}$ ($i,j$ mod 6), which is a 6-node clique from which all triangles are removed. Here the system increases the number of squares by disconnecting a fraction $\gamma$ of the nodes from the treelike bulk, to form  $\gamma N/6$ suitable disconnected 6-node graphlets. 

With the available numerical evidence at this moment, consisting of numerical solutions of order parameter equations and   spectra for ensembles with $\hat{\varrho}(\mu)=\alpha_3 \mu^3+\alpha_4 \mu^4$ (obtained via population dynamics with nontrivial acceptance probabilities), together with spectra measured in numerical graph simulations, satisfactory agreement has not yet been obtained. 
Preliminary population dynamics computations suggest that the above phase separation phenomena are not yet captured.  
Yet the structure of the theory seems elegant and intuitive, with the richness to exhibit complex phase transition phenomenology, 
and it is correct in the limit $\hat{\varrho}\to 0$.  The reason for the residual disagreement has not yet been identified. Apart from mundane explanations (e.g. imprecise or nonequilibrated numerical algorithms, either in population dynamics or graph generation, or saddle-points that are not of the assumed form), a possible candidate lies in the usage of complex logarithms. Some of the identities used in the present derivations, such as $\log \exp(Z)=Z$ or $\log(ZW)=\log(Z)+\log(W)$,  hold for complex values only if we can steer away from the cut in the complex plane. A more careful re-derivation may well reveal extra terms that for some reason become irrelevant for $\hat{\varrho}\to 0$ (where we know the theory works) but may be important for $\hat{\rho}\neq 0$.  To this author, the structure of the equations simply feels right as a description of loopy graph ensembles, and he is optimistic that the remaining discrepancy will soon be removed. 

\subsection{Processes on loopy graphs}

Looking ahead towards the use of the presently proposed method for the modelling of spin systems on loopy graphs, one can see that the calculation of the disorder-averaged free energy density of such systems will involve further (non-complex) replicas, and requires the evaluation of the following more complicated generating function:
\begin{eqnarray}
\Phi_K[\hat{\varrho},\{\bsigma\}]&=& \frac{1}{N}\log \sum_{\bc}\rme^{N\int\!\rmd\mu~\hat{\varrho}(\mu)\varrho(\mu|\bc)+K\sum_{i<j}c_{ij}\bsigma_i\cdot\bsigma_j}
\prod_{i\leq N}\delta_{k_i,\sum_j c_{ij}}
\end{eqnarray}
with $K=\beta J$ and the $n$-replicated spin vectors $\bsigma_i=(\sigma_i^1,\ldots,\sigma_i^n)$. Again also this quantity can be calculated, the limit $N\to\infty$ can be taken, and one finds a more complex closed set of equations in which now the graph order parameters and the spin order parameters are entangled. If we assume spin replica symmetry, in addition to graph replica symmetry, the final RS order parameter will take the form $\WW[\{x,y\},v]$, in which $v$ denotes an effective field. 
It is encouraging to confirm that upon assuming the physical saddle-point to be of the form $\WW[\{x,y\},v]=\WW[\{x,y\}]\WW[v]$ (i.e. the statistical features of the process decouple from those of the graph) we indeed recover the previous equations of this paper for $\WW[\{x,y\}]$, and the solution of \cite{Bethe} for the transition temperature on tree-like graphs (where indeed one would expect such decoupling to apply).

\section*{Acknowledgements}

The author gratefully acknowledges valuable discussions with 
Alessia Annibale, Reimer K\"{u}hn, Federico Ricci-Tersenghi, and Pierpaolo Vivo.

\end{document}